\documentclass[aps, prd, nofootinbib, preprintnumbers, showpacs]{revtex4}
\usepackage{graphicx}
\usepackage{amsmath}
\usepackage{multirow}
\usepackage{bm}
\usepackage{color}
\def\fsl#1{\setbox0=\hbox{$#1$}                 % set a box for #1
   \dimen0=\wd0                                 % and get its size
   \setbox1=\hbox{/} \dimen1=\wd1               % get size of /
   \ifdim\dimen0>\dimen1                        % #1 is bigger
      \rlap{\hbox to \dimen0{\hfil/\hfil}}      % so center / in box
      #1                                        % and print #1
   \else                                        % / is bigger
      \rlap{\hbox to \dimen1{\hfil$#1$\hfil}}   % so center #1
      /                                         % and print /
   \fi}                                         %

\newcommand{\VEV}[1]{\langle #1 \rangle}

\begin{document}
\title{Enhanced diphoton Higgs decay rate and isospin symmetric Higgs boson}
\date{\today}
\author{Michio Hashimoto}
 \email{michioh@isc.chubu.ac.jp}
  \affiliation{
   Chubu University, \\
   1200 Matsumoto-cho, Kasugai-shi, \\
   Aichi, 487-8501, JAPAN}
 \author{V.A. Miransky}
   \email{vmiransk@uwo.ca}
 \affiliation{
 Department of Applied Mathematics, \\
 Western University, \\
 London, Ontario N6A 5B7, CANADA}

\begin{abstract}
The ATLAS and CMS experiments have recently discovered a new $125$ GeV boson.
We show that the properties of this particle, including the enhancement of 
its diphoton decay rate,
can be explained in a model with an isospin 
symmetric Higgs boson. 
The predictions of the model relevant for future experiments are 
also discussed.
\end{abstract}

\maketitle

\section{Introduction}

Recently, the ATLAS \cite{atlas} and CMS \cite{CMS} experiments 
at the Large Hadron Collider (LHC) reported that a new boson $h$, 
compatible to the Standard model (SM) Higgs boson $H$,
was discovered in the mass range 125--126~GeV. On the other hand,
the ATLAS and CMS data might already suggest existence of
a new physics beyond the SM: While the decay channels of 
$h \to ZZ^{*}$ and $h \to WW^{*}$
are fairly consistent with the SM, the diphoton branching ratio 
Br$(h \to \gamma \gamma)$ is about 1.6 times larger than 
the SM value\footnote{
To the contrary, Plehn and Rauch \cite{1207.6108} have recently argued that 
none of the measured couplings deviates from its SM values 
significantly.
Also, the QCD uncertainties are discussed in Ref.~\cite{1207.1451}.
Thus, the observed deviations are not yet definitive.
}. 
This deviation from the SM has been discussed by 
many authors~\cite{enh-diphoton}.

In this paper, 
we will show that the ATLAS and CMS data for 
the enhanced diphoton branching ratio can be explained 
in the class of models with isospin symmetric (IS) electroweak Higgs boson
suggested by the authors in Refs.~\cite{Hashimoto:2009xi,Hashimoto:2009ty}. 
It is noticeable that as will be shown below, 
these models also make several predictions, which can be checked at
the LHC in the near future.

\section{IS Higgs Models}
\label{2}

There is a large hierarchy between quark masses from different 
families \cite{pdg}. Besides, the isospin violation in different
families is also hierarchical. It is very strong in the third
family, strong (although essentially weaker) in the second family, and
mild in the first one:
$\frac{m_t}{m_b} \simeq 41.5$, 
$\frac{m_c}{m_s} \simeq 13.4$, and
$\frac{m_u}{m_d} \simeq 0.38\mbox{--}0.58$~\cite{pdg}.
This is a big mystery:
In the framework of the SM, it is required to introduce hierarchical 
Yukawa couplings by hand, e.g., $\frac{y_t^{\rm SM}}{y_b^{\rm SM}} \simeq 41.5$,
and $\frac{y_c^{\rm SM}}{y_s^{\rm SM}} \simeq 13.4$.

A class of models (the IS Higgs models) 
describing the hierarchies in the quark mass spectrum 
was previously studied in Refs. \cite{Hashimoto:2009xi,Hashimoto:2009ty}.
One of our main motivations for introducing such models
was to find a dynamical mechanism that could shed light on the experimental
fact that the isospin violation in the quark mass spectrum is 
essentially stronger in heavier families.

The main characteristics of these models are 
the following: 
(a) It is assumed that the dynamics primarily responsible for 
electroweak symmetry breaking (EWSB) leads to the mass 
spectrum of quarks with no (or weak) isospin violation.
{\it Moreover, it is assumed that the values of these masses are of 
the order of the observed masses of the down-type quarks.} 
(b) The second (central) assumption is introducing the horizontal 
interactions for the quarks in the three families. 
As a first step, a {\it subcritical} (although nearcritical, i.e., strong) 
diagonal horizontal interactions for the top quark is utilized 
which lead to the observed ratio $\frac{m_t}{m_b} \simeq 41.5$.
The second step is introducing {\it equal} strength (i.e., isospin
symmetric) horizontal flavor-changing-neutral (FCN) interactions between 
the $t$ and $c$ quarks and the $b$ and $s$ ones. 

All together, these interactions naturally provide the observed ratio 
$m_c/m_s \simeq 13.4$ in the second family~\cite{Hashimoto:2009xi}. 
As emphasized in Ref.~\cite{Hashimoto:2009xi}, the choice of 
the IS masses being close to the values of the observed masses of 
the down-type quarks is crucial in this scenario.
As to the mild isospin violation in the first family, 
it was studied together with the effects of the family mixing, 
reflected in the Cabibbo-Kobayashi-Maskawa (CKM) matrix~\cite{Hashimoto:2009xi}
(see also Sec. \ref{5} below).

In this scenario, besides the EWSB interactions, the dominant
dynamics responsible for the form of the mass spectrum of quarks is
connected with the diagonal horizontal interactions for the third family and
the horizontal, isospin symmetric, FCN interactions between 
the second and third ones.
One of the signatures of this scenario is the appearance of a
composite top-Higgs doublet $\Phi_{h_t}$ (resonance) composed of 
the quarks and antiquarks of the third 
family~\cite{Hashimoto:2009xi,Hashimoto:2009ty}\footnote{
Such composites in the nearcritical regime in a symmetric phase of 
models with dynamical chiral symmetry breaking
were studied by several authors~\cite{Chivukula:1990bc}.}.

Thus, the main source of
the isospin violation in this approach is only the strong top quark 
interactions. 
On the other hand, because these interactions are subcritical, 
the top quark plays a minor role in EWSB.
The latter distinguishes this scenario from the top quark condensate 
model~\cite{Miransky:1988xi,Nambu,Marciano:1989xd,Bardeen:1989ds,Hashimoto:2000uk,Hill:2002ap}.
Note that unlike the topcolor assisted technicolor model
(TC2)~\cite{Hill:1994hp}, this class of models utilizes subcritical
dynamics for the top quark, so that without strong fine tuning, the
bosons from the top-Higgs doublet $\Phi_{h_t}$ are heavy, say, of order 1~TeV, 
in general (compare with Ref. \cite{Hashimoto:2009ty}).

Although the concrete model in Refs.~\cite{Hashimoto:2009xi,Hashimoto:2009ty}
utilized the fourth family of fermions \cite{He:2001tp,Frampton}
for generating EWSB, this choice is not crucial, as the authors emphasized
in \cite{Hashimoto:2009xi}.
In particular, the fourth family can be replaced by just a IS Higgs
boson doublet $\Phi_{h}$, without specifying its composite origin (if any).
In this paper, we will consider just such a 
version in which 
the neutral scalar from the $\Phi_{h}$ doublet will be identified with 
the 125~GeV $h$ boson.
Here we emphasize that while the neutral top-Higgs boson $h_t$ has
a large top-Yukawa coupling,
the IS Higgs boson $h$ does not, $y_t \simeq y_b \sim 10^{-2}$.
On the other hand, the $hWW^{*}$ and $hZZ^{*}$ coupling constants are
close to those in the SM (see below). Also, the mixing between
$\Phi_{h}$ and much heavier $\Phi_{h_t}$ should be small 
(compare with \cite{Hashimoto:2009ty}).
Let us now describe the decay processes of the IS Higgs $h$.

\section{Decay modes $h \to \gamma\gamma$, 
$h \to Z\gamma$, $h \to WW^*$, and $h \to ZZ^*$}
\label{3}

Let us consider the diphoton branching ratio in the IS Higgs model.
It is well known that the $W$-loop contribution to $H \to \gamma\gamma$
is dominant in the SM, while the top-loop effect is destructive
against the $W$-loop. More concretely, the diphoton partial width in 
the SM reads~\cite{diphoton}: 
\begin{equation}
  \Gamma (H \to \gamma\gamma) = \frac{\sqrt{2} G_F \alpha^2 m_H^3}{256\pi^3}
  \bigg| A_1 (\tau_W) + N_c Q_t^2 A_{\frac{1}{2}} (\tau_t)\bigg|^2,
  \qquad
  \tau_W \equiv \frac{m_H^2}{4m_W^2}, 
  \tau_t \equiv \frac{m_H^2}{4m_t^2}, 
\end{equation}
where $G_F$ denotes the Fermi constant,
$N_c=3$ represents the number of colors, and $Q_t=+2/3$ is 
the electric charge of the top quark.
The loop functions $A_1$ and $A_{\frac{1}{2}}$ for $W$ and $t$, 
respectively, are given by
\begin{equation}
  A_1 (\tau) \equiv - \frac{1}{\tau^2} 
  \bigg[2 \tau^2 + 3\tau + 3(2\tau-1)f(\tau)\bigg],
\end{equation}
and 
\begin{equation}
  A_{\frac{1}{2}}(\tau) \equiv \frac{2}{\tau^2} 
  \bigg[\tau + (\tau -1) f(\tau)\bigg] , 
\end{equation}
with $f(\tau) \equiv \arcsin^2 \sqrt{\tau}$ for $\tau \leq 1$. 
Then, the numerical values of the $W$- and $t$-loop functions read
\begin{equation}
  A_1 (\tau_W) = -8.32, \qquad A_{\frac{1}{2}}(\tau_t) = 1.38,
\end{equation}
for $m_W = 80.385$ GeV~\cite{pdg}, $m_t = 173.5$ GeV~\cite{pdg}, 
and $m_H = 125$ GeV.

On the other hand,
in the IS Higgs model, the Yukawa coupling between the top and 
the IS Higgs $h$ is as small as the bottom Yukawa coupling,
so that the top-loop contribution is strongly suppressed.
The partial decay width of $h \to \gamma\gamma$ is thus enhanced 
without changing essentially $h \to ZZ^{*}$ and $h \to WW^{*}$.
A rough estimate taking the isospin symmetric 
top and bottom Yukawa couplings $y_t \simeq y_b \approx 10^{-2}$
is as follows:
\begin{equation}
\label{gamma}  
  \frac{\Gamma^{\rm IS} (h \to \gamma \gamma)}
       {\Gamma^{\rm SM}(H \to \gamma \gamma)} \simeq 1.56 , \qquad
  \frac{\Gamma^{\rm IS} (h \to WW^{*})}
       {\Gamma^{\rm SM}(H \to WW^{*})} = 
  \frac{\Gamma^{\rm IS} (h \to ZZ^{*})}
       {\Gamma^{\rm SM}(H \to ZZ^{*})} = 
       \left(\frac{v_h}{v}\right)^2 \simeq 0.96 . 
\end{equation}
Here using the Pagels-Stokar formula \cite{PS}, we estimated 
the vacuum expectation value (VEV) of the top-Higgs $h_t$ 
as $v_t = 50$ GeV, and
the VEV $v_h$ of the IS Higgs $h$ is given by the relation $v^2 = v_h^2 + v_t^2$
with $v = 246$ GeV. Note that
the values of the ratios in Eq. (\ref{gamma}) are not very sensitive to 
the value of $v_t$, e.g., for $v_t=40$--$100$ GeV,
the suppression factor in the pair decay modes to $WW^*$ and $ZZ^*$
is $0.97$--$0.84$ and 
the enhancement factor in the diphoton channel is $1.58$--$1.37$. 
For the decay mode of $h \to Z\gamma$, this model yields 
\begin{equation}
\label{Zgamma}
  \frac{\Gamma^{\rm IS} (h \to Z \gamma)}
       {\Gamma^{\rm SM}(H \to Z \gamma)} \simeq 1.07 
\end{equation}
(the data concerning this decay channel has not yet been
reported~\cite{atlas,CMS}).
Note that the total decay width is almost unchanged,
so that Eqs.~(\ref{gamma}) and (\ref{Zgamma}) indicate 
the suppression/enhancement factors of the corresponding branching ratios. 

The values in Eq. (\ref{gamma}) agree well with the data in 
the ATLAS and CMS experiments.
However, obviously, the main production mechanism of the Higgs boson,
the gluon fusion process $gg \to h$, is now in trouble.
The presence of new chargeless colored particles, which considered by
several authors \cite{adj-S}
can help to resolve this problem.
We pursue this possibility below. 

\section{Model with colored scalar}
\label{4}

We utilize an effective theory near the EWSB scale.
The model contains: 
(1) the IS Higgs doublet $\Phi_h$, 
which is mainly responsible for the EWSB and 
couples to the top and bottom in the isospin symmetric way, 
(2) the top-Higgs doublet $\Phi_{h_t}$, 
which is required to obtain the correct top mass, and
(3) the colored scalar and/or fermions 
which are required to enhance $gg \to h$. 

The items (1) and (2) above are essentially described in 
Refs.~\cite{Hashimoto:2009xi,Hashimoto:2009ty}. 
The only difference is that the 
two composite Higgs doublets composed of the fourth family quarks 
should now be replaced by the IS Higgs.
We will discuss this point later.
Note that in this case the Lagrangian density ${\cal L}$ in 
the effective theory contains the IS Higgs quartic coupling $\lambda_h$, 
${\cal L} \supset -\lambda_h |\Phi_h^\dagger \Phi_h|^2$, 
and the mass $m_h = 125$~GeV corresponds to a small $\lambda_h$ via
the relation $m_h^2 \simeq 2 \lambda_h v_h^2$ like in the SM, 
because the mixing between $\Phi_h$ and the much heavier $\Phi_{h_t}$ 
is tiny in the present model  
(compare with Refs. \cite{Hashimoto:2009xi,Hashimoto:2009ty}).
However, unlike the case of the SM~\cite{RGE-lam}, this does not imply that 
the theory keeps the perturbative nature up to some extremely 
high energy scale, as we will see below.

As to a concrete realization of item (3),
we may introduce a real scalar field $S$ in the adjoint
representation of the color $SU(3)_c$ and utilize the Higgs-portal 
model~\cite{adj-S}, just as a benchmark case,
\begin{eqnarray}
  {\cal L} \supset 
  {\cal L}_S = \frac{1}{2} (D_\mu S)^2 - \frac{1}{2} m_{0,S}^2 S^2 
  - \frac{\lambda_S}{4} S^4
  - \frac{\lambda_{h S}}{2} S^2 \Phi_h^\dagger \Phi_h,
  \qquad \Phi_h = \left(\begin{array}{c}
    \omega^+ \\ \frac{1}{\sqrt{2}}(v_h + h + i z_0)
  \end{array} \right),
  \label{Lag-S}
\end{eqnarray}
where $\omega^\pm$ and $z_0$ are the components eaten by $W^\pm$ and $Z$.
The scalar field $S$ is chosen to be assigned to 
the $({\bf 8},{\bf 1})_0$ representation 
of the $SU(3)_c \times SU(2)_W \times U(1)_Y$.
Other representations, for example, a color triplet, 
are also possible.
Note that we do not incorporate a Higgs-portal term  
between $S$ and $\Phi_{h_t}$ and 
other possible cubic and quartic terms into Eq.~(\ref{Lag-S}), 
because they do not play any important role in the following analysis.

The mass-squared term for the scalar $S$ is given by
\begin{equation}
  M_S^2 = m_{0,S}^2 + \frac{\lambda_{hS}}{2} v_h^2,
\end{equation}
and should be positive in order to avoid the color symmetry breaking.
Typically, $M_S \sim 200$~GeV is allowed in the current 
data~\cite{adj-S}.
We will take a positive value for $\lambda_{hS}$ 
and a classically (quasi-)scale invariant model with $m_{0,S}^2 \approx 0$, 
which is favorable to reproduce the SM like gluon fusion production.

Let us consider the contribution of the color octet $S$ to 
the gluon fusion process $gg \to h$ in the leading order,
\begin{equation}
  \frac{\sigma (gg \to h)}{\sigma^{\rm SM} (gg \to H)}
  \sim \frac{\Gamma(h \to gg)}{\Gamma^{\rm SM} (H \to gg)}
  = \bigg| \frac{C_A \lambda_{hS} \frac{v v_h}{2M_S^2} A_0(\tau_S)}
                {A_{\frac{1}{2}}(\tau_t)} \bigg|^2,
\end{equation}
with $C_A = 3$, $\tau_S \equiv m_h^2/(4 M_S^2)$, and 
\begin{eqnarray}
  A_0 (\tau) \equiv - \frac{1}{\tau^2} \bigg[\tau-f(\tau)\bigg] \, .  
\end{eqnarray}
We find $A_0 \simeq 0.37$--$0.34$ for $M_S=150$--$400$ GeV, so that
an appropriate value of the Higgs-portal coupling is
\begin{equation}
  \lambda_{hS} \simeq 2.5 \mbox{--} 2.7 \times \frac{M_S^2}{v v_h} \,.
\end{equation}
As a typical value, we may take $\lambda_{hS} = 1.8$ for 
$M_S= 200$~GeV and $v_t=50$~GeV.
When $m_{0,S}^2 \approx 0$, i.e., $M_S^2 \approx \lambda_{hS} v_h^2/2$,
we obtain $\Gamma(h \to gg) \approx 0.6 \times \Gamma^{\rm SM} (H \to gg)$,
independently of the values of $\lambda_{hS}$.
In order to stabilize the Higgs potential for $S$ at the tree level,
the relation $|\lambda_{hS}| < 2 \sqrt{\lambda_S \lambda_h}$ is 
also required.

A comment concerning the IS Higgs  quartic coupling $\lambda_h$ is in order.
In the SM, the Higgs mass 125~GeV suggests that the theory is 
perturbative up to an extremely high energy scale~\cite{RGE-lam}.
On the contrary, in the present model, when we take a large Higgs-portal 
coupling $\lambda_{hS}$ that reproduces $gg \to h$ correctly, 
the quartic coupling $\lambda_h$ will grow because
the $\beta$-function for $\lambda_h$ contains the $\lambda_{hS}^2$ term.
Also, there is no large negative contribution to 
the $\beta$-function for $\lambda_h$ from 
the top-Yukawa coupling $y_t \sim 10^{-2}$.

\begin{figure}[t]
  \begin{center}
  \includegraphics[width=7.5cm]{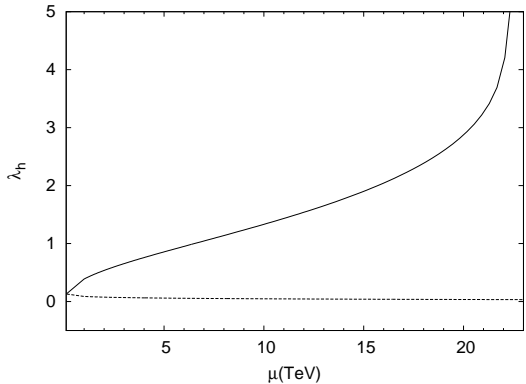}
   \end{center}
   \caption{The running behavior of the IS Higgs quartic coupling $\lambda_h$.
   The solid and dashed lines correspond to $\lambda_h$
   and the SM Higgs quartic coupling, respectively.
   We fixed the IS Higgs mass $m_h = \sqrt{2\lambda_h} v_h = 125$~GeV and
   took $\lambda_{hS}=1.8$ and $\lambda_S=1.5$.
   Unlike the SM, the IS Higgs quartic coupling grows up due to
   a large Higgs-portal coupling $\lambda_{hS}$  and a small
   top-Yukawa coupling $y_t$.
   \label{lam}}
\end{figure}

One can demonstrate such a behavior more explicitly by using the 
renormalization group equations.
In Fig.~\ref{lam}, the running of the coupling $\lambda_h$ is shown.
The IS Higgs mass is $m_h = \sqrt{2\lambda_h} v_h$, and we take it 
to be equal to $125$ GeV. 
Taking a large Higgs-portal coupling $\lambda_{hS}=1.8$ and 
the $S^4$-coupling $\lambda_S=1.5$, 
it turns out that the coupling $\lambda_h$ rapidly grows.
Due to the running effects, 
the naive instability of the scalar potential at the tree level 
is resolved around the TeV scale in this case.
The blowup scale strongly depends on the initial values of $\lambda_{hS}$ 
and $\lambda_S$.
A detailed analysis will be performed elsewhere.
Last but not least, we would like to mention that 
other realizations of the enhancement of the $h$ production 
are also possible.

\section{Quark mass matrices}
\label{5}

Let us discuss the structure of the quark mass matrices in the present model.
The Yukawa interactions are written by~\cite{Hashimoto:2009xi,Hashimoto:2009ty} 
\begin{equation}
  - {\cal L}_Y =
    \sum_{i,j}\bar{\psi}_L^{(i)} Y_D^{ij} d_R^{(j)} \Phi_{h}
  + \sum_{i,j}\bar{\psi}_L^{(i)} Y_U^{ij} u_R^{(j)} \tilde{\Phi}_{h}
  + y_{h_t} \bar{\psi}_L^{(3)} t_R \tilde{\Phi}_{h_t},
\end{equation}
with
\begin{equation}
  \tilde{\Phi}_{h} \equiv i\tau_2 \Phi_h^*, \quad
  \tilde{\Phi}_{h_t} \equiv i\tau_2 \Phi_{h_t}^*, \quad
  \Phi_{h_t} = \left(\begin{array}{c}
    \omega_t^+ \\ \frac{1}{\sqrt{2}}(v_t + h_t + i z_t)
  \end{array} \right), \quad
\VEV{\Phi_h} = 
  \left(\begin{array}{c} 0 \\ \frac{v_h}{\sqrt{2}}\end{array}\right), \quad
  \VEV{\Phi_{h_t}} = 
  \left(\begin{array}{c} 0 \\ \frac{v_t}{\sqrt{2}}\end{array}\right), \quad
\end{equation}
\begin{equation}
  Y_D \equiv \frac{\sqrt{2}}{v_{h}} M_D, \quad
  Y_U \equiv \frac{\sqrt{2}}{v_{h}} M_U, 
\end{equation}
and
\begin{equation}
  M_D =
  \left(\begin{array}{ccc}
    m_0^{(1)} & \xi_{12} m_0^{(1)} & \xi_{13} m_0^{(1)} \\[2mm]
    \xi_{21} m_0^{(1)} & m_0^{(2)} + \delta \cdot m_b & \xi_{23} m_0^{(2)} \\[2mm]
    \xi_{31} m_0^{(1)} & \xi_{32} m_0^{(2)} & m_0^{(3)} \\[2mm]
  \end{array}\right) , \qquad
  M_U =
  \left(\begin{array}{ccc}
    \eta_{11} m_0^{(1)} & \eta_{12} m_0^{(1)} & \eta_{13} m_0^{(1)} \\[2mm]
    \eta_{21} m_0^{(1)} & m_0^{(2)} + \delta \cdot m_t & \eta_{23} m_0^{(2)} \\[2mm]
    \eta_{31} m_0^{(1)} & \eta_{32} m_0^{(2)} & m_0^{(3)} \\[2mm]
  \end{array}\right) ,  
\end{equation}
where $\psi_L^{(i)}$ denotes the weak doublet quarks from the $i$-th family,
and $u_R^{(i)}$ and $d_R^{(i)}$ represent the right-handed up- and down-type
quarks.
The top-Higgs part is responsible for the top mass, 
$m_t \simeq y_{h_t} \frac{v_t}{\sqrt{2}}$.
The IS masses $m_0^{(i)}$ are the same mass scales as the down-type quarks,
say, $m_0^{(3)} \sim \mbox{1 GeV}$, $m_0^{(2)} \sim \mbox{100 MeV}$, and 
$m_0^{(1)} \sim \mbox{1 MeV}$. 
The common one-loop factor $\delta \sim 1/100$ yields 
the correct mass hierarchy between $m_s$ and $m_c$ via the hierarchy
between $m_b$ and $m_t$.
Also, the off-diagonal coefficients are assumed to be
$\xi_{ij}, \eta_{ij} \sim {\cal O}(1)$, with some dynamical mechanism.
(We kept $\eta_{11}$ in the up sector for generality.)
The CKM matrix is approximately determined by the down-type quark mass
matrix~\cite{Hashimoto:2009ty},
\begin{equation}
  V_{\rm CKM} \approx
  \left(\begin{array}{ccc}
    1-\frac{|\xi_{12}|^2}{2}\left(\frac{m_0^{(1)}}{m_0^{(2)}}\right)^2 &
    \xi_{12} \frac{m_0^{(1)}}{m_0^{(2)}} & \xi_{13} \frac{m_0^{(1)}}{m_0^{(3)}} \\
   -\xi_{12}^* \frac{m_0^{(1)}}{m_0^{(2)}} &
    1-\frac{|\xi_{12}|^2}{2}\left(\frac{m_0^{(1)}}{m_0^{(2)}}\right)^2 &
    \xi_{23} \frac{m_0^{(2)}}{m_0^{(3)}} \\
   -(\xi_{13}^* - \xi_{12}^* \xi_{23}^*)\frac{m_0^{(1)}}{m_0^{(3)}} &
   -\xi_{23}^* \frac{m_0^{(2)}}{m_0^{(3)}} & 1 \\
  \end{array}\right) \, .
\end{equation}
We can then reproduce the CKM matrix, basically.
For example, with the inputs, 
$m_0^{(1)}=10$~MeV, $m_0^{(2)}=68$~MeV, $m_0^{(3)}=4.2$~GeV, $m_t=173.5$~GeV,
$\delta=7\times 10^{-3}$, 
$\xi_{12}=\xi_{21}=\eta_{12}=\eta_{21}=2.0$, 
$\xi_{13}=\xi_{31}=\eta_{13}=\eta_{31}=1.6$, 
$\xi_{23}=\xi_{32}=\eta_{23}=\eta_{32}=-2.5$, $\eta_{11}=\frac{1}{4}$, 
we obtain $m_d=4.9$~MeV, $m_s=95$~MeV, $m_b=4.2$~GeV, 
$m_u=2.2$~MeV, $m_c=1.3$~GeV, 
$|V_{ud}| \simeq |V_{cs}| = 0.975$, $|V_{tb}| \simeq 1$, 
$|V_{us}| \simeq |V_{cd}| = 0.22$, 
$|V_{cb}| = 0.041$, $|V_{ts}|=0.039$, $|V_{ub}| = 0.0042$, $|V_{td}| = 0.013$. 
These values fairly agree with the PDG ones~\cite{pdg}.

As was emphasized above in Sec. \ref{2}, because of the subcriticality dynamics 
in this class of models, the extra bosons
from the top Higgs doublet $\Phi_{h_t}$
are heavy, say, ${\cal O}(1 \mbox{TeV})$.
Thus, their one-loop contributions to the $B^0$-$\bar{B}^0$ mixing,
$b \to s \gamma$ and $Z \to b \bar{b}$ are suppressed.
A tree FCNC term also appears in the up sector,
so that the $D^0$--$\bar{D}^0$ mixing is potentially dangerous.
However, because the FCNC coupling $Y_{t-c-h_t}$ is found to be tiny, 
$Y_{t-c-h_t} \sim \frac{m_t}{v_t} \frac{m_u}{m_t} \frac{m_c}{m_t} =
10^{-6}\mbox{--}10^{-7}$, this does not cause any troubles.

\section{Conclusion}
\label{6}

The model with an IS Higgs boson yields not only an explanation
of the ATLAS and CMS data, including the enhanced diphoton Higgs decay rate, 
but also makes several predictions. The most important of them is that
the value of the top-Yukawa coupling $h$-$t$-$\bar{t}$ should be close 
to the bottom-Yukawa one.
Another prediction relates to the decay mode $h \to Z\gamma$, 
which unlike $h \to \gamma\gamma$ is enhanced only slightly, 
$\Gamma^{\rm IS} (h \to Z \gamma) = 1.07 \times \Gamma^{\rm SM} (H \to Z\gamma)$.
Last but not least, the LHC might potentially discover 
the top-Higgs resonance $h_t$, if lucky.

\acknowledgments 
This work is supported by the Natural Sciences and 
Engineering Research Council of Canada.

\end{document}